\documentclass{iaus}
\usepackage{graphicx}

\newcommand{\farcs}{\mbox{\ensuremath{.\!\!^{\prime\prime}}}}%

\title[Populations of OB stars in galaxies] 
{Populations of OB-type stars in galaxies}

\author[C.~J.~Evans]   
{Christopher J. Evans$^1$}

\affiliation{$^1$ UK Astronomy Technology Centre, Blackford Hill, Edinburgh, 
EH9 3HJ, UK \\ email: {\tt chris.evans@stfc.ac.uk}
}

\pubyear{2011}
\volume{272}  
\pagerange{xxx--xxx}
\jname{Active OB Stars}
\editors{C. Neiner, G. Wade, G. Meynet \& G. Peters, eds.}
\begin{document}

\maketitle

\begin{abstract}
One of the challenges for stellar astrophysics is to reach the point
at which we can undertake reliable spectral synthesis of unresolved
populations in young, star-forming galaxies at high redshift.  Here I
summarise recent studies of massive stars in the Galaxy and Magellanic
Clouds, which span a range of metallicities commensurate with those in
high-redshift systems, thus providing an excellent laboratory in which
to study the role of environment on stellar evolution.  I also give an
overview of observations of luminous supergiants in external galaxies
out to a remarkable 6.7\,Mpc, in which we can exploit our understanding of stellar
evolution to study the chemistry and dynamics of the host systems.

\end{abstract}

\firstsection 
\section{Introduction}
One of the prime motivations to study stellar evolution is to use
that knowledge to develop tools to explain integrated-light 
observations of distant star clusters and galaxies.
Consider the recent multi-wavelength study of a
gravitationally-lensed galaxy at a redshift of $z$\,$=$\,2.3 by
\cite{ams10}. By virtue of the lens, individual regions
are resolved in sub-millimetre imaging, each $\sim$100\,pc in
scale. These are intense regions of star formation on a comparable
spatial scale to that of 30~Doradus, viewed at a time when the
universe was significantly younger.  Such observations are only
possible at present due to the magnification of the lens but, with
future facilities such as the Atacama Large Millimetre Array (ALMA)
and Extremely Large Telescopes (ELTs), we can expect comparable
observations in unlensed systems in the coming years.
One of the real tests of stellar astrophysics is to reach the point
at which we are confident that we can interpret integrated-light
spectroscopy of such distant systems accurately, exploiting our
understanding of massive stars to obtain new insights into the
processes at work during one of the most critical epochs of galaxy
evolution.

\begin{figure}[h]
\begin{center}
\includegraphics[width=12.5cm]{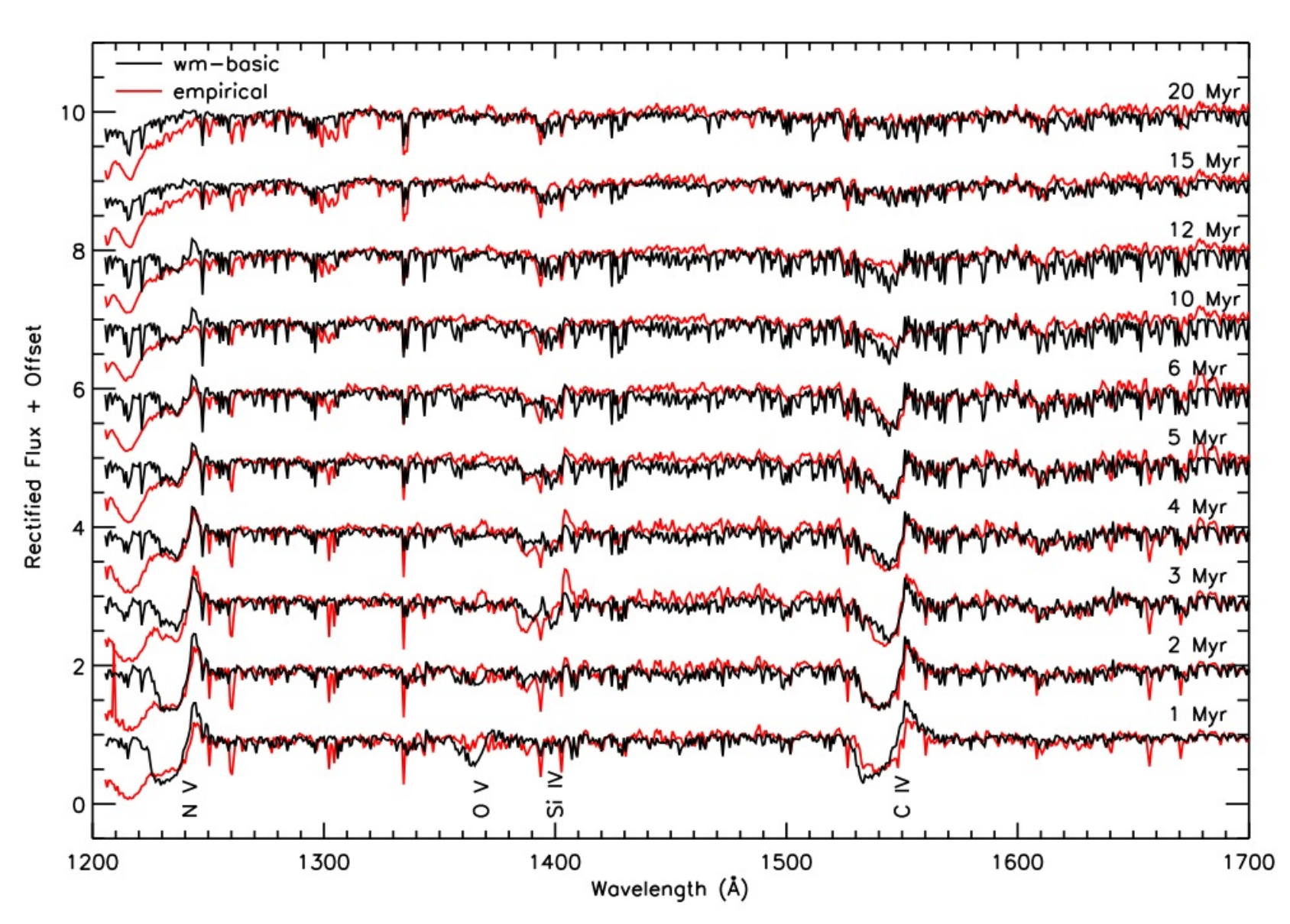} 
\caption{Single-stellar populations from the new Starburst99
WM-Basic library (solar metallicity, black lines) compared
to empirical {\em IUE} templates (red lines), from Leitherer 
et al. (2010). \label{fig1}}
\end{center}
\end{figure}

Population synthesis codes such as Starburst99 (Leitherer et al. 1999)
are the `bridge' from studies of individual stars to analysis of
entire populations on galaxy scales.  The rest-frame
ultraviolet (UV) is rich with the signatures of stellar winds and, for
high-redshift galaxies, is redshifted into the optical (e.g.  Pettini
et al. 2002) and, ultimately, into the near-infrared for the most
distant systems.  To model the rest-frame UV a new spectral library
for Starburst99 has been calculated by Leitherer et al. (2010) using the
WM-Basic model atmosphere code (Pauldrach, Hoffmann \& Lennon, 2001).
Fig.~\ref{fig1} shows a comparison of library spectra from the {\em
International Ultraviolet Explorer (IUE)} with spectra calculated
using the new WM-Basic library.  In general there is excellent agreement,
a worthy testament to theoretical developments in recent years!  Many of the apparent differences are related to
observational issues such as narrow interstellar features and the
significant wings of the Lyman-$\alpha$ absorption from
Galactic~H\,{\scriptsize I}.  The most discrepant stellar feature is
O\,{\scriptsize V} $\lambda$1371\AA\/, known to be sensitive to
wind inhomogeneities (`clumping'), the effects of which are not
incorporated in WM-Basic at present.

The comparison in Fig.~\ref{fig1} reminds me of comment that a
high-redshift astronomer once said to me at a conference: {\em `...but
stars are done aren't they?'}  At the time I argued strongly to the contrary
-- while we now have a significant command of stellar astrophysics, a
broad range of fundamental questions still eludes us for massive
stars, including basic issues such as their formation (Zinnecker \&
Yorke, 2007) and end-points (Smith et al. 2010).

We can use observations of populations of massive stars to
improve our understanding of the physics and evolution of the stars
themselves and, once we are confident of a decent grasp of their behaviour,
we can use them as tracers of the properties of their host galaxies.
In the following sections, I summarise recent developments in terms 
of the role of environment on the evolution of
massive stars (Sec.~\ref{sec2}), observational studies of luminous
supergiants in external galaxies (Sec.~\ref{sec3}), and 
the potential of the next generation of ground-based
telescopes in studies of massive stars (Sec.~\ref{sec4}).

\section{Massive Stars \& Metallicity}\label{sec2}

The Large and Small Magellanic Clouds (LMC and SMC) are metal
deficient when compared to the solar neighbourhood, with metallicities
($Z$) of approximately 50\% and 20-25\% solar (e.g. Trundle et
al. 2007). This presents us with the chance to study the
properties of massive stars in regimes which span a
range in metallicity comparable to those found in galaxies in the
early universe (e.g. Erb et al. 2006; Nesvadba et al. 2008).

In the course of investigating near-IR photometry from the {\em
Spitzer Space Telescope}, Bonanos et al. (2009, 2010, these
proceedings) have compiled catalogues of published spectral
classifications for over 7,000 massive stars in the Clouds.  This
illustrates the significant progress we have made in determining the
stellar content of two of our nearest neighbours.  Over the past
decade the focus has been to exploit the latest generation of
ground-based telescopes to deliver sufficient spectral resolution and
signal-to-noise for quantitative atmospheric analysis of large samples
of massive stars in the Clouds.  This includes the VLT-FLAMES Survey
of Massive Stars (Evans et al. 2005, 2006), and more targeted studies
of O-type stars (Massey et al. 2004, 2005, 2009) and B-/Be-type stars
(Martayan et al. 2006, 2007).  Analysis of these samples has been used
to investigate a broad range of parameters, including the
$Z$-dependence of stellar wind intensities, effective temperatures,
and rotational velocities, each of which is now briefly discussed.

\subsection{Metallicity-dependent stellar winds}

\begin{figure}[h]
\begin{center}
\includegraphics[width=4.5in]{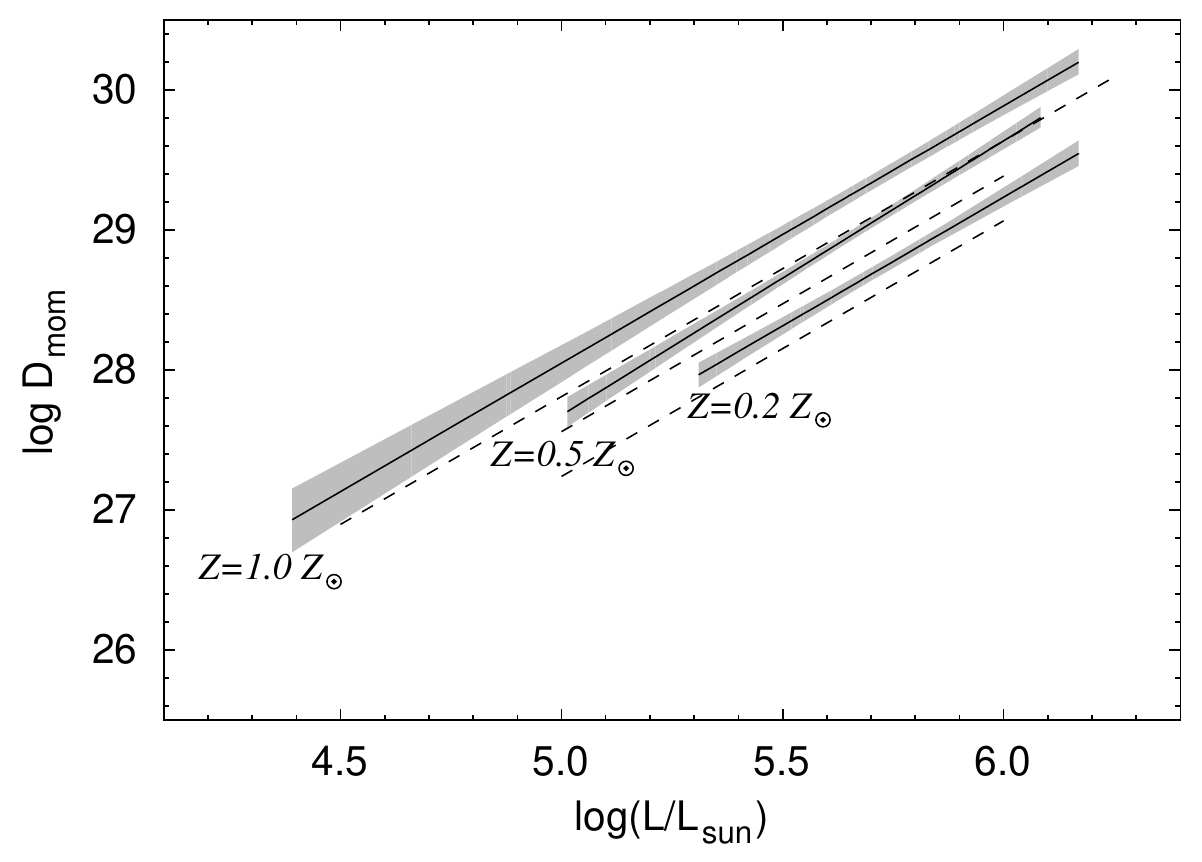} 
\caption{The observed wind-momentum--luminosity relations (solid lines) 
for O-type stars from the FLAMES survey (Mokiem et
al. 2007b) compared with theoretical predictions (dashed lines).  The
upper, middle and lower relations are the Galactic, LMC and SMC
results, respectively. \label{fig11}}
\end{center}
\end{figure}

A consequence of the theory of radiatively-driven stellar winds is
that their intensities should be dependent on $Z$, with weaker winds
at lower metallicities (Kudritzki et al. 1987; Vink et al. 2001).
Indications for such an offset were seen from analysis of 22 stars in
the Clouds (Massey et al. 2005), with more comprehensive evidence
provided by the larger sample from the FLAMES survey (Mokiem et al.
2007b). The FLAMES results are shown in Fig.~\ref{fig11}, with fits to the stellar
wind-momenta, D$_{\rm mom}$ (which is a function of the mass-loss rate,
terminal velocity and stellar radius).  Fig.~\ref{fig11} also shows
the theoretical predictions using the prescription from \cite{vink01}
-- the relative separations of the observed fits are in good
agreement, finding a $Z$-dependence with exponents in the range
0.72-0.83 (depending on assumptions regarding clumping in the winds),
as compared to $Z^{0.69\pm0.10}$ from theory.  Analysis of early
B-type supergiants in the SMC (Trundle et al. 2004; Trundle \& Lennon
2005) also reveals weaker winds compared to their Galactic
counterparts (Crowther, Lennon \& Walborn, 2006).

\subsection{Stellar Effective Temperatures}

The effects of line-blanketing are reduced at lower metallicities due
to the diminished cumulative opacity from the metal lines. This leads
to less `back warming' by trapped radiation, thus requiring a hotter
model to reproduce the observed line ratios.

The consequences of this are clearly seen in the temperatures obtained
for O-type stars in the SMC, which are hotter than those found for
Galactic stars with the same spectral type (Massey et al. 2005; Mokiem
et al. 2006).  The temperatures for stars in the LMC are seen to fall
neatly between the SMC and Galactic results (Mokiem et al.  2007a).  A
similar $Z$-dependence was also found in the effective temperatures
derived for B-type stars observed with FLAMES (Trundle et al. 2007).
Typical differences from this effect are 5-10\% when moving from solar
to SMC metallicity. Note that there are also
temperature differences for cooler supergiants (e.g. Evans \& Howarth,
2003; Levesque et al. 2006), but in the opposite sense and for
different reasons (see Evans, 2009).

\subsection{Stellar rotational velocities}

Stellar rotation strongly affects the evolution of all O- and
B-type stars via, for example, changes in their main-sequence
lifetimes (Meynet \& Maeder, 2000). The effects of rotation also
manifest themselves via mixing of chemically-processed material,
leading to changes in the surface abundances of elements such as
nitrogen (e.g. Brott et al., these proceedings).

The increased fraction of Be- to normal B-type stars at lower
metallicity (Maeder, Grebel \& Mermilliod, 1999) pointed to a
$Z$-dependence in stellar rotation rates. Investigations into the
effects of $Z$ on stellar rotational rates have largely focussed on
studies of B-type stars, to avoid the potential complications of
angular momentum loss due to strong stellar winds in more massive
stars.  In the Magellanic Clouds it can be difficult to define an
appropriate `field' or `cluster' sample for comparison, given that the
rotation rates for stars in clusters appear faster than for those in
the field population (Keller, 2004; Strom et al. 2005; Wolff et
al. 2008). However, a trend of faster velocities at lower $Z$ appears
to be borne out by comparisons between rotation rates in the Galaxy
and the LMC (Keller, 2004) and, more recently, including the SMC
(Martayan et al. 2007; Hunter et al. 2008).

Mokiem et al. (2006) found tentative evidence for different rotational
velocity distributions for O-type stars in the SMC compared to the
Galaxy, but with some reservations given the potential effects of
mass-loss and macroturbulence. New results suggest that
macroturbulence is ubiquitous in O-type stars (e.g. Sim\'{o}n-D\'{i}az
et al., these proceedings), with Penny \& Gies (2009) suggesting that
its magnitude is also $Z$-dependent.  This perhaps points to an origin
similar to the convective effects argued by Cantiello et al. (2009) to
account for {\em microturbulence} in massive stars (also seen to have
a $Z$-dependence).

\subsection{What next?}

The VLT-FLAMES Tarantula Survey is a new ESO Large Programme
which has obtained multi-epoch spectroscopy of over 1,000 stars
in the 30~Doradus region of the LMC (Evans et al. 2010a).  30~Dor is
the largest H\,{\scriptsize II} region in the Local Group, providing us
with an excellent stellar nursery to build-up a large observational
sample of the most massive stars; the top-level motivations for the
survey are summarised by Lennon et al. (these proceedings).

Multi-epoch, radial velocity studies of OB-type stars in Galactic
clusters have found binary fractions in excess of 50\% (see Sana
\& Evans, these proceedings), and the results of Bosch, Terlevich
\& Terlevich (2009) suggest a similarly large binary fraction for the
O-type stars in 30~Dor.  The Tarantula Survey was designed to combine
high-quality spectroscopy for quantitative atmospheric analysis, with
repeat observations for detection of massive binaries (over a longest
baseline of one year).

This strategy has already paid dividends in terms of putting strong
constraints on the nature of 30~Dor\,\#016, a massive O2-type star on
the western fringes of 30~Dor (Evans et al. 2010b).  Previous
spectroscopy revealed a peculiar radial velocity, but the new FLAMES
spectra enabled a massive companion to be ruled out to a high level of
confidence, suggesting the star as an ejected runaway.  

A further example from the new survey is the observation of two
separate components in some of the spectra of R139 (Taylor et al., in
preparation).  R139 is just over 1$'$ to the north of R136, the dense
cluster at the core of 30~Dor, and was classified by Walborn \& Blades
(1997) as O7~Iafp.  Previously reported as a
single-lined binary with a period of 52.7\,d (Moffat, 1989),
recent efforts by Schnurr et al. (2008) found no evidence for
binarity in R139 (within a range of periods of up to 200\,d), although
they noted it as having a slightly variable radial velocity.
Chen\'{e} et al. (these proceedings) also report new observations of
radial velocity variations in R139 from a separate monitoring
campaign.

\section{Beyond the Magellanic Clouds}\label{sec3}

The BA-type supergiants are the most intrinsically luminous (`normal')
stars at optical wavelengths.  The 8-10\,m class telescopes have given
us the means by which we can obtain spectroscopy for quantitative
analysis of individual blue supergiants in galaxies well beyond the
Magellanic clouds, providing us with estimates of their physical
parameters, chemical abundances, and also providing an alternative
distance diagnostic in the form of the flux-weighted
gravity--luminosity relationship (Kudritzki, Bresolin \& Przybilla,
2003; Kudritzki et al. 2008).  

Analysis of high-resolution spectroscopy of luminous A-type
supergiants in each of NGC\,6822 (Venn et al. 2001), M31 (McCarthy et
al. 1997, Venn et al. 2000), WLM (Venn et al. 2003) and Sextans~A
(Kaufer et al. 2004) provided some of the first examples of this type
of work in external galaxies, with analysis of larger samples of B-type
supergiants in M31 by Trundle et al. (2002) and in M33 by Urbaneja
et al. (2005b)

More recently, the Araucaria project has combined optical and near-IR
imaging of Cepheids with spectroscopy of luminous blue supergiants to
refine the distance determinations to nearby galaxies (Gieren et
al. 2005).  The Araucaria project has obtained low resolution
($\sim$5\,\AA) optical spectroscopy with the VLT of BA-type
supergiants in a number of external galaxies, as summarised
in Table~\ref{tab1}.

\begin{table}[!h]
\begin{center}
\caption{Summary of published spectroscopic observations and analyses of luminous 
blue supergiants in external galaxies from the Araucaria project. \label{tab1}}
{\scriptsize
\begin{tabular}{lccl}\hline
{\bf Galaxy} & {\bf d [Mpc]} & {\bf 12$+$log(O/H)} & {\bf References} \\
\hline
IC\,1613 & 0.7 & 7.90\,$\pm$\,0.08 & Bresolin et al. (2007) \\
WLM & 0.9 & 7.83\,$\pm$\,0.12 & Bresolin et al. (2006), Urbaneja et al. (2008) \\
NGC\,3109 & 1.3 & 7.76\,$\pm$\,0.07 & Evans et al. (2007) \\
NGC\,300 & 1.9 & $-$ & Bresolin et al. (2002); Urbaneja et al. (2003); Urbaneja et al. (2005a) \\
NGC\,55 & 1.9 & $-$ & Castro et al. (2008) \\
\hline
\end{tabular}
}
\end{center}
\end{table}

This has provided us with estimates of oxygen abundances/metallicities
in the irregular dwarf galaxies IC\,1613 and WLM, and in the main body of
NGC\,3109. In the larger spirals, the stellar
abundances allow study of the radial abundance trends (e.g. Urbaneja
et al. 2005a), providing a useful comparison for 
nebular abundance determinations.
In the case of NGC\,300, Bresolin et al. (2009) find
that `strong-line' nebular diagnostics (such as the $R_{23}$ ratio)
overestimate the true abundance by a factor of two (or more) compared
to estimates from auroral nebular lines and the stellar analyses, 
highlighting the caution that should be employed in metallicity estimates
in distant star-forming galaxies (e.g. Kewley \& Ellison, 2008).

\section{Future Prospects in the Era of ELTs}\label{sec4}

Observations from the Araucaria project highlight the power of using
individual stars to investigate the properties of their host galaxies.
Indeed, one of the most remarkable observations in this context is the
VLT spectroscopy of two luminous supergiants in NGC\,3621, at a
distance of 6.7\,Mpc (Bresolin et al. 2001).  However, with the
exception of the very brightest stars, our potential of exploring the
massive star content of external galaxies is limited by the sensitivity
of current facilities.  With primary apertures in excess of 20\,m, the 
ELTs, the next generation of optical-IR telescopes, will revolutionise
our ground-based capabilities, particularly when coupled with adaptive
optics (AO) to correct for atmospheric turbulence, thus delivering
huge gains in both sensitivity and angular resolution.  

Spectroscopy of stellar populations in external galaxies is one of the
key elements of the science cases toward the ELTs.  This includes
using stars as tracers of properties of the host galaxies, and to
extend our studies of environmental effects on stellar evolution in
systems such as starbursts and very metal-poor galaxies.
Efforts toward building the ELTs are increasingly global, with
three projects now in the advanced stages of their design,
fund-raising and planning for instrumentation -- the Giant Magellan
Telescope (GMT), the Thirty Meter Telescope (TMT), and the European
Extremely Large Telescope (E-ELT).

\subsection{MAD: An AO pathfinder for ELTs}\label{mad}

An integral part of the ELTs is the use of AO to deliver improved
image quality in the near-IR, and there is an understandable desire to
maximise the field-of-view over which one can obtain both good and
uniform correction.  A key technical development in plans
toward the E-ELT was an on-sky demonstration of multi-conjugate
adaptive optics (MCAO), which uses multiple deformable mirrors to
correct for different layers of turbulence in the atmosphere.  This
was realised as the Multi-conjugate Adaptive optics Demonstrator
(Marchetti et al. 2007), delivering AO-corrected near-IR imaging over
a 1\,$\times$\,1\,arcmin field. MAD was commissioned at the VLT in
early 2007.

\begin{figure}[h]
\centering
\includegraphics[scale=0.45]{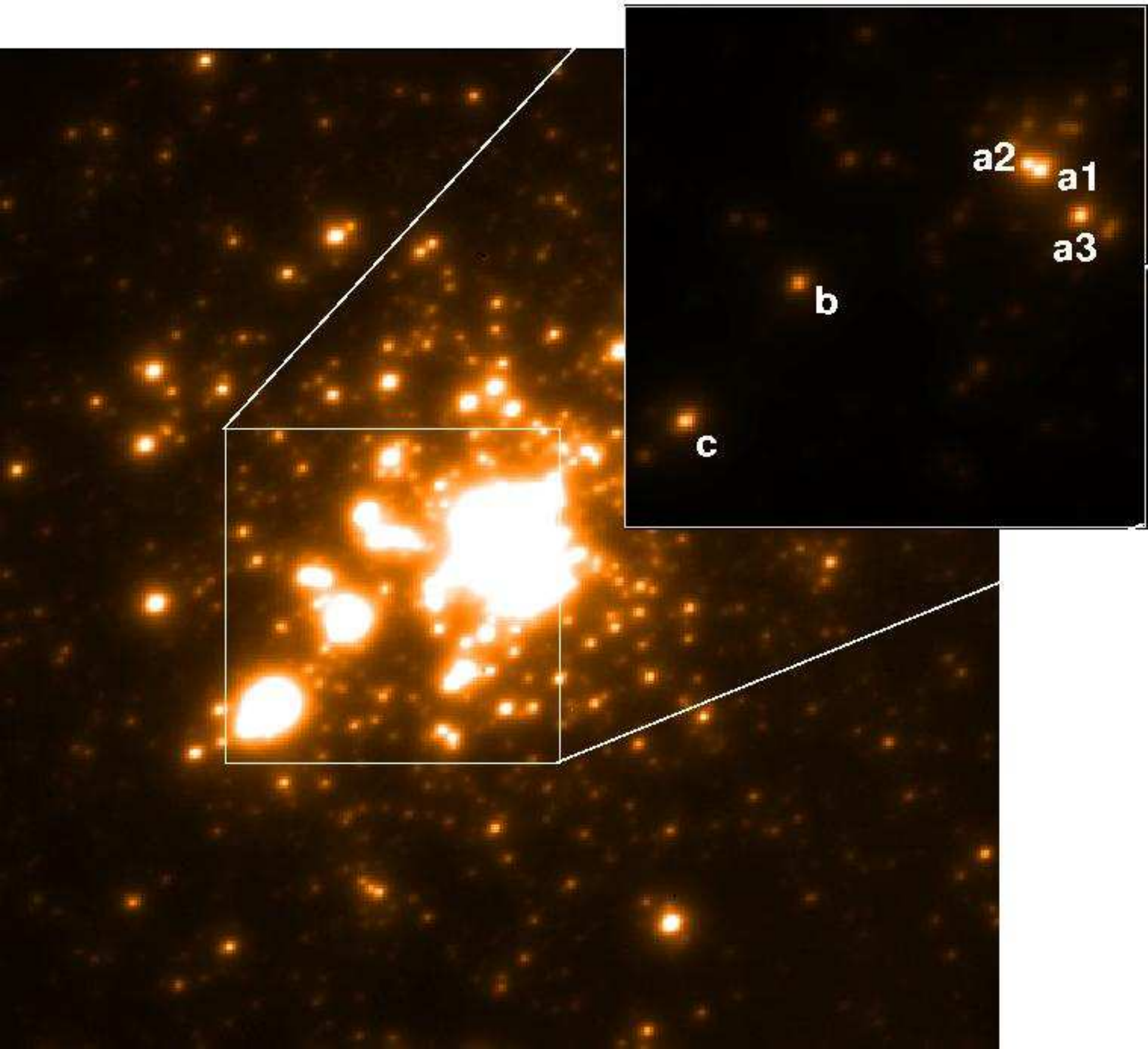}
\caption{Central 12$''$\,$\times$\,12$''$ (eqv. 3\,$\times$\,3\,pc) of the $K_{\rm s}$-band
MAD image, with an expanded view of the central 4$''$\,$\times$\,4$''$ showing 
the four WN5h very massive stars analysed by Crowther et al. (2010; 
components a1, a2, a3, and c).\label{madpac}}
\end{figure}

As part of a science demonstration programme, MAD was used to obtain
$H$- and $K_{\rm s}$-band imaging of R136 (Campbell et al. 2010).  The
exquisite resolution achieved in these data is shown by
Fig.~\ref{madpac}, in which the two central WN5h stars in the core of
R136 (`a1' and `a2', separated by 0\farcs1) are spatially resolved.
This provides near-comparable angular resolution to optical imaging
with the {\em HST} (e.g. Hunter et al. 1995), at wavelengths less
affected by the significant and variable extinction toward 30~Dor.

When combined with AO-corrected IFU spectroscopy from Schnurr et
al. (2009), these data have been used by Crowther et al. (2010) to
argue that three of the central stars in R136 have current masses in
excess of 150\,M$_{\odot}$.  From cluster IMF simulations they find
the massive stars in R136 are consistent with an upper initial
mass-limit of $\sim$300\,M$_{\odot}$. This contrasts with the claim by
Figer (2005) of an upper limit of 150\,M$_{\odot}$ from analysis of
the Arches cluster.  To address this difference, Crowther et
al. returned to the most massive members of the Arches with new
photometry and revised distance/extinction estimates, finding greater
current masses than those from Martins et al. (2008), and consistent
with evolutionary tracks for stars with initial masses in excess of
150\,M$_{\odot}$.

\section{Summary}

Work continues apace in characterising the massive star populations in
external galaxies (e.g. Herrero et al. 2010) and to improve our
understanding of stellar evolution in massive O-type stars (e.g. the
Tarantula Survey).  We now have a handle on the role of metallicity on
some key aspects of stellar evolution, but there remain a number of
issues to be addressed before we can claim to really understand
`populations' in distant galaxies, such as the effects of binary
evolution and revision to the accepted upper mass limit for single
stars.

Inclusion for the effects of binary evolution has been argued to
account for some of the observed properties of low-redshift galaxies
(e.g. Han, Podsiadlowski \& Lynas-Gray, 2007; Brinchmann, Kunth \&
Durret, 2008). The high incidence of massive binaries (Sana \&
Evans, these proceedings) suggests further work is required to
characterise their typical properties and assess their impact on our
interpretation of unresolved populations.

Lastly, the new results of Crowther et al. (2010) urge for
investigation of the effects of very massive stars in
population synthesis codes.  Although very rare, and almost certainly
short-lived, Crowther et al. show that the four WN5h stars at
the core account for $\sim$45\% of the Lyman-continuum
ioinizing flux and $\sim$35\% of the mechanical power of the respective 
totals from stars in the central 5\,pc of R136.  The inferred star-formation rates for the
components of high-redshift galaxies from Swinbank et al. (2010) and
Jones et al. (2010) are an order of magnitude larger than that from
30~Dor but one could, very naively, image scaling-up the density and intensity of
R136 to larger scales -- in which case the initial mass function would suggest
a considerable input of radiation/energy from very massive stars.

\acknowledgements{I gratefully acknowledge financial support from the IAU.}

\end{document}